\documentclass{IEEEcsmag}
\usepackage{amsmath,amsfonts}
\usepackage{algorithmic}
\usepackage{algorithm}
\usepackage{array}
\usepackage[caption=false,font=normalsize,labelfont=sf,textfont=sf]{subfig}
\usepackage{textcomp}
\usepackage{stfloats}
\usepackage{url}
\usepackage{verbatim}
\usepackage{graphicx}
\usepackage[numbers]{natbib}
\usepackage{graphics}
\usepackage{adjustbox}  
\usepackage{tabularx}  
\usepackage{makecell}
\usepackage{longtable}
\usepackage{tablefootnote}
\usepackage{afterpage}
\usepackage[flushleft]{threeparttable}
\usepackage{orcidlink}
\usepackage{upmath}

\begin{document}

\title{RIS-Vis: A Novel Visualization Platform for Seismic, Geodetic, and Weather Data Relevant to Antarctic Cryosphere Science}

\author{Aishwarya Chakravarthy \orcidlink{0009-0003-4481-2844}}
\affil{Georgia Institute of Technology, Atlanta, GA, 30332, USA}

\author{Dhiman Mondal \orcidlink{0000-0003-0708-6575}}
\affil{Massachusetts Institute of Technology, Haystack Observatory, Westford, MA, 01886, USA}

\author{John Barrett \orcidlink{0000-0002-9290-0764}}
\affil{Massachusetts Institute of Technology, Haystack Observatory, Westford, MA, 01886, USA}

\author{Chet Ruszczyk \orcidlink{0000-0001-7278-9707}}
\affil{Massachusetts Institute of Technology, Haystack Observatory, Westford, MA, 01886, USA}

\author{Pedro Elosegui \orcidlink{0000-0002-4120-7855}}
\affil{Massachusetts Institute of Technology, Haystack Observatory, Westford, MA, 01886, USA}

\begin{abstract}
\looseness-1
Antarctic ice shelves play a vital role in preserving the health of the Antarctic cryosphere and beyond. By serving as a buttressing force, ice shelves prevent sea-level rise by restraining the flow of continental ice and glaciers to sea. Sea-level rise impacts the global environment in multiple ways, including flooding habitats, eroding coastlines, and contaminating groundwater. It is therefore essential to monitor the stability of Antarctic ice shelves, for which various data sources are required. We have developed RIS-Vis, a novel data visualization platform to monitor Antarctic ice shelves. RIS-Vis is capable of analyzing and visualizing seismic, geodetic, and weather data to provide meaningful information for cryosphere research. RIS-Vis was built using Python libraries including Obspy, APScheduler, and the Plotly Dash framework, and uses SQLite as the backing database. Visualizations developed on RIS-Vis include seismic waveforms, spectrograms, power spectral densities, and graphs of geodetic ice-shelf flow and meteorological variables.

\end{abstract}

\maketitle

\section{Introduction}
\chapteri{I}ce shelves play an integral role in preserving the health of the Antarctic cryosphere system by moderating the flow of grounded ice from the massive ice sheet that covers the continent to the ocean. Over recent years, the collapse of several Antarctic ice shelves including the Larsen-B ice shelf has led to the acceleration of glaciers into the ocean, contributing to sea-level rise \cite[e.g.,][]{ref1, ref2}. In fact, it has been estimated that the combined ice-shelf mass loss from the East Antarctic Ice Sheet (EAIS) and the West Antarctica Ice Sheet (WAIS) has contributed to global sea-level rise an average of $4.6\pm 1.2\,\mathrm{mm}$ between 1992 and 2017 \cite{ref3}. Thus, an observing system to monitor changes in the state of balance of Antarctic ice shelves continuously and over large ranges of time would prove to be an extremely useful tool for glaciologists, oceanographers, and scientists in general.

We are exploring new \textit{in situ} approaches for monitoring Antarctic ice-shelf stability that involve a new observing instrument and an accompanying dashboard system. The instrument, called the Seismo-Geodetic Ice Penetrator (SGIP), is equipped with geophysics-grade sensors such as a broadband seismometer and a geodetic-quality GPS receiver. Once completed, a network of SGIP instruments will be air-dropped onto the Ross Ice Shelf (RIS), where they will continuously collect seismic and geodetic data. The data will be transferred to the Internet in near real-time via satellite communications to then be analyzed by scientists. 
\indent In this paper, we present RIS-Vis, an Antarctic ice-shelf data visualization system capable of supporting the SGIP instruments by providing data visualization capabilities for scientists. RIS-Vis helps researchers monitor and visualize near real-time seismic, geodetic, weather, and engineering (i.e., instrument subsystems) data that can be collected either from freely available repositories or by a monitoring instrument such as SGIP. In the following sections, after briefly reviewing related work, we present the RIS-Vis data sources, design considerations, key features, development, and challenges, and close by discussing possible future directions and enhancements.

\section{Related Work}
Web visualizations are a popular tool utilized by data scientists to make conclusions based on large datasets. However, there has been less emphasis on applying popular data visualization capabilities to track environmental variables and conditions such as the state of Antarctic ice shelves.

\textbf{Antarctic Visualizations.} Prior research has made efforts to visualize temporal changes in sea ice extent around Antarctica. For example, the National Snow and Ice Data Center (NSIDC) publishes interactive line charts and image visualizations of Antarctic sea-ice extent that are updated daily \cite{ref4}. Researchers have also built an augmented reality platform to visualize ice-shelf structure using data from the ROSETTA-Ice project, which thoroughly mapped the RIS from 2015 to 2018 \cite{ref5}. The visualization platform was built for Microsoft HoloLens and allows users to interact with Lidar and radar data \cite{ref5}. In addition, NASA Scientific Visualization Studio (SVS) develops several visualizations relevant to Antarctic cryosphere science. These include visualizations for tracking ice-height change and ice-mass loss compiled from several satellite missions \cite{ref6}. However, the SVS visualizations are not updated in near real-time.

In general, although various organizations track and publish visualizations of Antarctic ice-shelf data in various fashions, a singular web visualization platform that consolidates many aspects of the state of the Antarctic ice shelves, making the science comprehensive and accessible in near real-time, has been lacking.

\begin{table*}[!t]  
    \caption{Summary of RIS-Vis Data Sources}
    \fontsize{3pt}{4pt}\selectfont
    \centering
    \begin{adjustbox}{width=\textwidth}  
    \begin{threeparttable}
        \begin{tabular}{|c|c|c|}
            \hline
            \textbf{Data Type} & \textbf{Database Name} & \textbf{Station Name (GPS Coordinates)}\textsuperscript{\textdagger} \\
            \hline
            Seismic & \makecell{Incorporated Research Institutions for Seismology (IRIS) \\Data Management Center (DMC)} & 
            \makecell{HOO (-77.531, 166.93)\\CONZ (-77.532, 167.08) \\ELHT (-77.510, 167.14) \\NAUS (-77.521,167.15)}\\
            \hline
            Geodetic & Nevada Geodetic Laboratory & \makecell{
            FTP4 (-78.928,162.56) \\
            MTCX (-78.520,162.53) \\
            LMRG (-78.097,163.85)\\
            DNTH (-78.079,163.98)
            }\\
            \hline
            \makecell{Weather (Long-term weather data \\stored in database)} & \makecell{University of Wisconsin-Madison \\Automatic Weather Station (AWS) Network} & \makecell{
            MARILYN (-79.897,165.85)}\\
            \hline
            \makecell{Weather (Real-time weather data \\NOT stored in database)} & \makecell{OpenWeatherMap API} & \makecell{
            N/A (-81.500,-175.00)}
            \\
            \hline
            System Monitoring & Proxy Data & N/A (N/A) \\
            \hline
        \end{tabular}
        \begin{tablenotes}
            \item\textsuperscript{\textdagger}See Figure~\ref{fig:data_map}
         \end{tablenotes}
    \end{threeparttable}
    \end{adjustbox}
    \label{table:data_sources}
\end{table*}

\section{Data Sources}
The first SGIP systems are expected to be deployed on the RIS towards the end of 2024 (i.e., during the Antarctic 2024 summer field season). SGIP data will begin to become available in near real-time soon after, that is after a brief instrument commissioning period following deployment. To exercise SGIP data visualization in advance, RIS-Vis currently incorporates SGIP-like data from existing sensors from a variety of publicly available sources. These include data from the Incorporated Research Institutions for Seismology (IRIS) Data Management Center (DMC) \textemdash{}a seismic data repository \cite{ref7}\textemdash{}, the Nevada Geodetic Laboratory \textemdash{}a geodetic data repository \cite{ref8}\textemdash{}, and the University of Wisconsin-Madison Antarctic Automatic Weather Station (AWS) Project \textemdash{}a weather data repository \cite{ref9}. Table~\ref{table:data_sources} provides a summary of the data sources and their usage within RIS-Vis. Although these data repositories are not equipped to be updated in real-time, RIS-Vis was developed with scalability in mind to account for the near real-time data transfers from the SGIP. 

Currently, RIS-Vis incorporates four main types of data, including:

\textbf{Seismic Data.} Ice shelves experience continuous forcings from a variety of environmental stressors, such as the ocean and the atmosphere, causing vibrations, flexure, deformation, crevassing, melting, and other effects, ultimately driving them to calving, or making them susceptible to collapse \cite{ref10}. For example, infragravity waves are long-period ocean gravity waves that can cause ice shelves to flex, and thus have the potential to induce stresses that can worsen existing crevasses and compromise ice-shelf integrity \cite{ref11}. Seismic data can be used to probe the response of ice shelves to ocean waves, thus monitoring the conditions that might lead to their collapse. For that reason, the SGIP science payload includes a broadband seismometer.

\textbf{Geodetic Data.} Ice shelves also flow, drift, and deform at rates that are typically much slower than those associated with seismic frequencies. Geodetic measurements using Global Navigation Satellite System (GNSS; i.e., GPS and also other constellations such as GLONASS and Galileo) can track the changing positions of glacier locations with high precision \cite[e.g.,][]{child2021,schild2021}. Monitoring the positions of an ice shelf before, during, and after iceberg calving and collapse would help scientists improve their understanding of the processes that lead to those events. Consequently, the SGIP payload includes a geodetic-quality GNSS receiving system.

\textbf{Weather Data.} Global warming has significantly impacted Antarctic climate; in fact, measurements indicate that surface temperature in Western Antarctica has increased at a rate more than twice the global average \cite{ref12, ref13}. Meteorological data such as temperature, pressure, humidity, and wind provide the atmospheric information necessary to assist with the processing and interpretation of seismogeodetic results. Continuous long-term monitoring of ice-shelf weather would help scientists determine the impact of climate change on the state of the ice shelves and the effect on sea level rise. The SGIP instrument includes a meteorological package to measure the conditions in the close vicinity of the instrument. 

\textbf{System Monitoring Data.} Engineering data such as sensor voltages and currents, battery capacity, and communication throughput are critical for the understanding of the state, health, and performance of the SGIP instrument. RIS-Vis has already developed this capability but currently utilizes proxy data because proper system-monitoring data from SGIP will not be available until the instruments are deployed on the RIS.

\begin{figure*}
\centering
\includegraphics[width=6in]{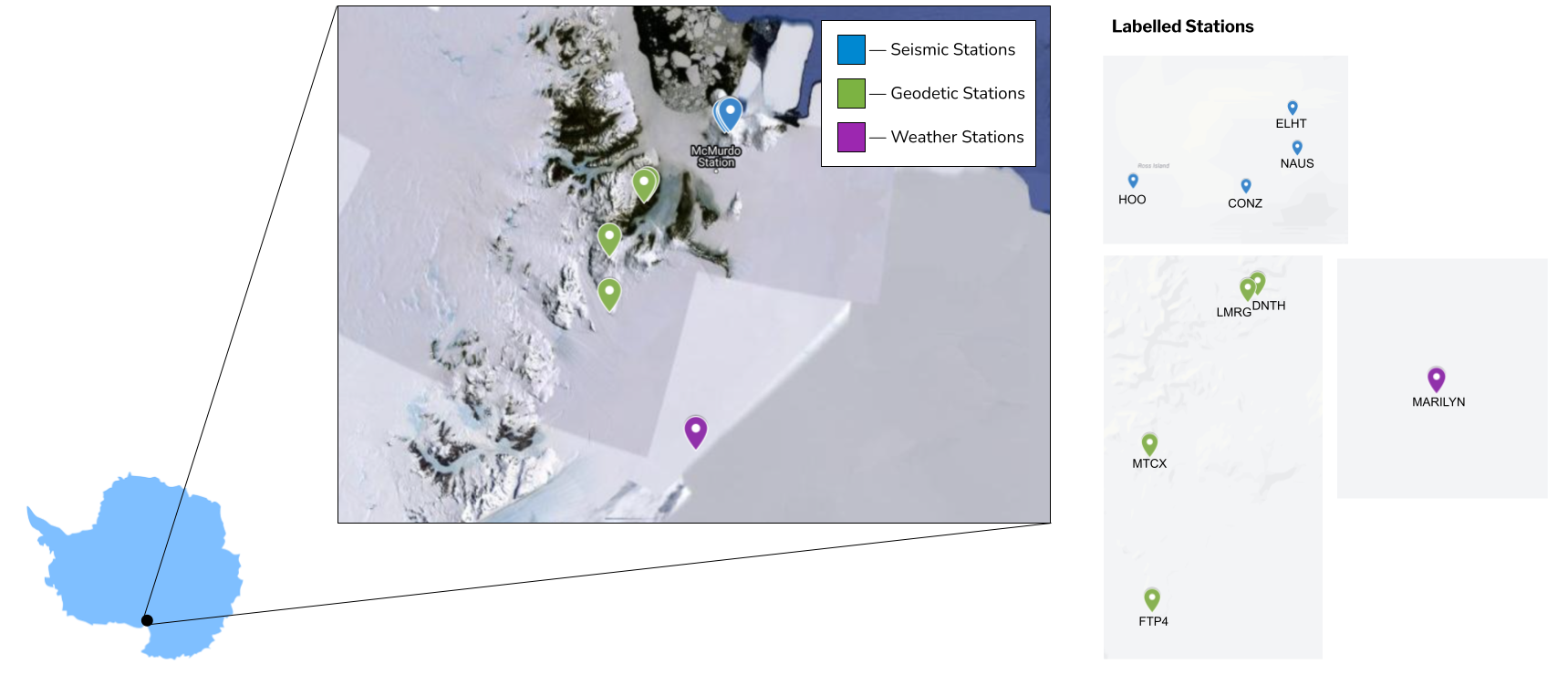}
\caption{Proxy seismic, geodetic, and weather station locations in the vicinity of McMurdo Base, Antarctica, used in this study (see also Table~\ref{table:data_sources}).}
\label{fig:data_map}
\end{figure*}

\section{Design Considerations}
We considered three main factors when architecting and building RIS-Vis:
    
\begin{itemize}
    \item \textbf{Incorporating multiple sources of data.} As discussed above, monitoring the Antarctic ice-shelf state involves many factors that oftentimes significantly influence each other. Although there have been efforts in the past to study individual aspects of Antarctic ice shelf weather, vibrations, or flow \cite[e.g.,][]{ref11}, to the best of our knowledge, there has not been a singular system built to handle monitoring multiple stressors relevant to ice shelf stability. Creating a monitoring dashboard capable of processing seismic, geodetic, and weather data (and other data types that may become available in the future) in near-real time allows scientists to examine independent data with an interconnected lens, and to analyze different data types within the same time period.
    
    \item \textbf{Creating interactive visualizations.} Scientists often create visualizations to better extract key findings from large datasets. With respect to Antarctic ice shelf health, few softwares have so far been created to display visualizations of seismic, geodetic, and weather data in an easy-to-use and interactive format. RIS-Vis addresses this challenge by displaying visualizations in a web dashboard application, in which data processing is abstracted. In addition, as RIS-Vis displays data using Plotly Dash (see below), a web framework designed specifically for creating interactive dashboards, users are able to examine the data in a more hands-on manner.
    
    \item \textbf{Real-time monitoring and long-term data storage.} There are two main needs when developing an ice shelf health monitoring application: data should be processed in near-real time and data should be available long-term for visualization. RIS-Vis handles these needs by scheduling downloads from data sources in real-time and storing data for long-term visibility in a database. In addition, within the web application, RIS-Vis offers capabilities both for real-time monitoring of the Ross Ice Shelf and for examining data from the far past. These features allow researchers to be able to analyze current developments in the Ross Ice Shelf without delays, as well as to analyze long-term changes in the RIS with ease.
\end{itemize}

\section{Key Features of RIS-Vis}
\urlstyle{tt}
The three key features that best define the RIS-Vis visualization platform are:

\begin{itemize}
\item{\textbf{Open-source web dashboard} tool that helps scientists examine seismic, geodetic, and weather data from the Ross Ice Shelf. (RIS-Vis is currently hosted on Github at \url{https://github.com/AishwaryaC26/RIS-Vis}; documentation is provided on the website.)}

\item{\textbf{Interactive visualization design} that displays graphs in which data can be zoomed in on, panned over, selected, and more using visualization features of Plotly Dash.}

\item{\textbf{Scalable implementation} in which both real-time data and long-term data can be visualized. For example, weather and geodetic visualizations are functional with selections spanning multiple years.}
\end{itemize}

\section{RIS-Vis Development}

We next describe in detail the architecture of RIS-Vis including its back-end, front-end, and some optimization decisions made.

\begin{figure*}[!t]
\centering
\includegraphics[width=7in]{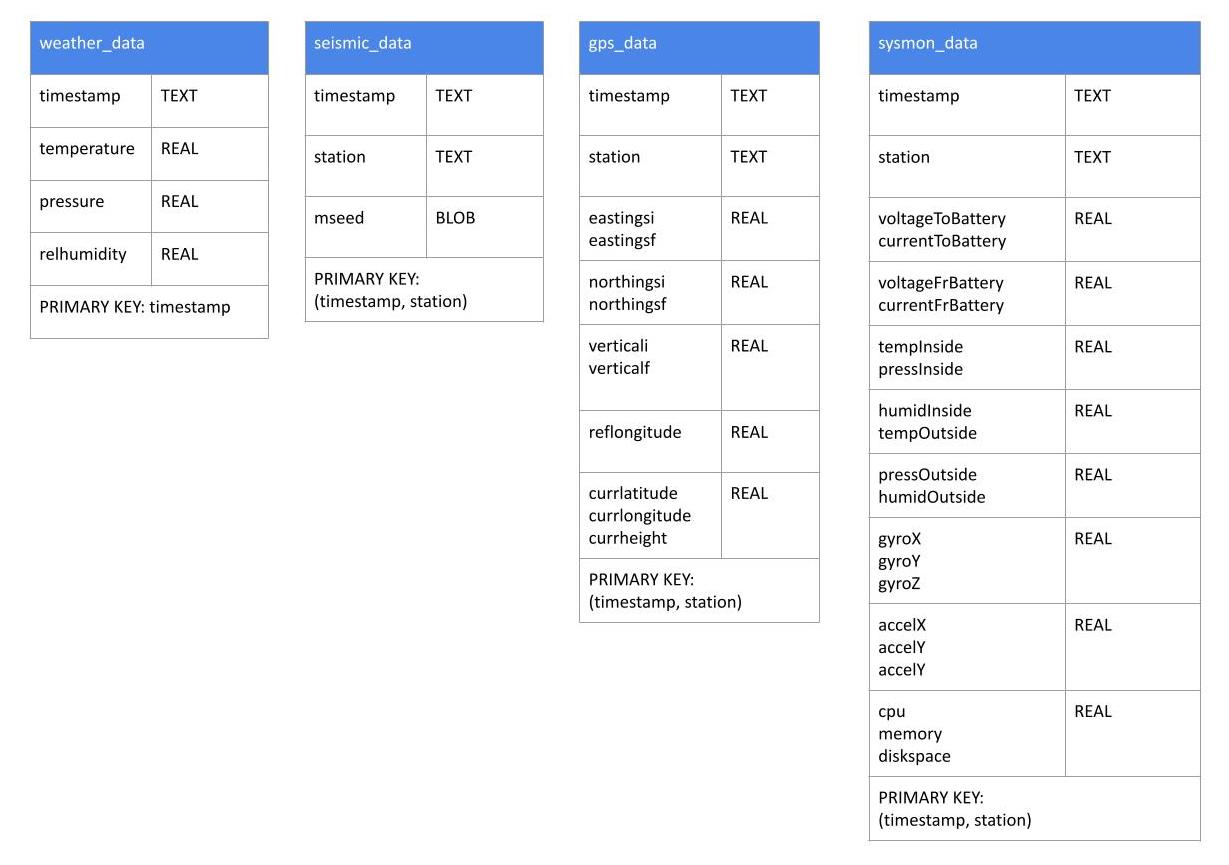}%
\hfil
\caption{RIS-Vis SQLite database schema. Similar variables are combined into one row (e.g., eastingsi, eastingsf).}
\label{fig:sqlite_schema}
\end{figure*}

\subsection{Back-end} 

We chose SQLite as the backing database for RIS-Vis due to its simplicity and efficiency. SQLite databases are stored as a single file and are efficient, particularly for time series data. As RIS-Vis does not utilize concurrent writing operations (i.e., inserting multiple data into the database at the same time), the major limitations of SQLite do not affect the performance of RIS-Vis.

The schema of the SQLite database implementation of RIS-Vis is detailed in Figure~\ref{fig:sqlite_schema}. This schema is able to accommodate the addition of new stations, if necessary, due to the incorporation of a “Station” column within the appropriate tables. Worthy of note is the fact that seismic data are stored as BLOBs within the SQL database. This is because the IRIS DMC commonly stores and returns data as miniSEED files, which is a compressed version of SEED (the most common form of storing seismological data).

RIS-Vis facilitates real-time downloading of seismic, geodetic, and weather data from various online repositories. To do so, RIS-Vis utilizes Python’s APScheduler library. APScheduler allows the scheduling of specific functions. For example, in RIS-Vis, seismic data and GPS data were scheduled to be downloaded once daily to approximately match typical uploads to those data repositories, but weather data was scheduled to be downloaded once monthly due to the delay in data uploads to the University of Wisconsin-Madison weather database. (Seismogeodetic data downloading from the SGIP instruments is expected to occur hourly.)

\subsection{Front-End}

\subsubsection{Libraries used}

The front-end of RIS-Vis was developed using a variety of open-source Python libraries, capable of both supporting the dashboard user interface as well as the data analysis procedures. The open-source libraries significantly used as well as their purpose within the project include:
    \begin{itemize}
        \item{\textbf{Plotly Dash}
        \begin{list}{}{}
            \item{\textbf{Description:} Web framework capable of building interactive visualization dashboards (built on top of Python's Flask web framework)\cite{ref14}.}
            \item{\textbf{Purpose:} Utilized as web framework for RIS-Vis.}
        \end{list}
        }
        \item{\textbf{Advanced Python Scheduler}
        \begin{list}{}{}
            \item{\textbf{Description:} Task scheduler and queue system \cite{ref15}.}
            \item{\textbf{Purpose:} Utilized to schedule downloads of seismic, GPS, and weather data.}
        \end{list}
        }
        \item{\textbf{Obspy}
        \begin{list}{}{}
            \item{\textbf{Description:} Framework for downloading and processing seismological data \cite{ref16}.}
            \item{\textbf{Purpose:} Utilized to download and process seismic data from the Ross Ice Shelf.}
        \end{list}
        }
    \end{itemize}
    
    Other data analysis and visualization libraries used include MatPlotlib, Numpy, and Pandas.\\

\setcounter{figure}{3}
\begin{figure*}[htbp] 
    \centering 
    \subfloat[]{\includegraphics[width=5in]{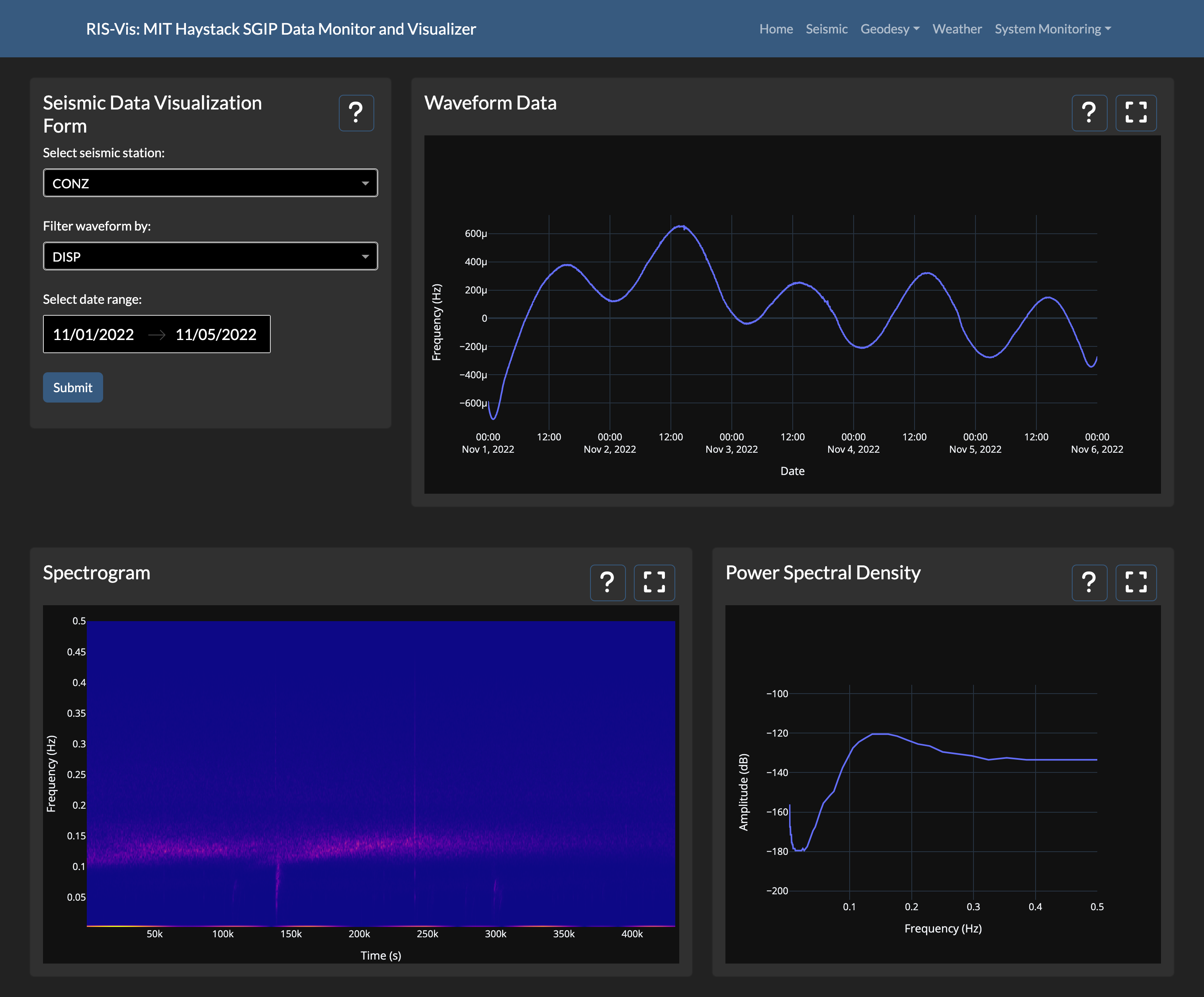}%
    \label{fig:seismic_page}}
    \hfil
    \centering 
    \subfloat[]{\includegraphics[width=5in]{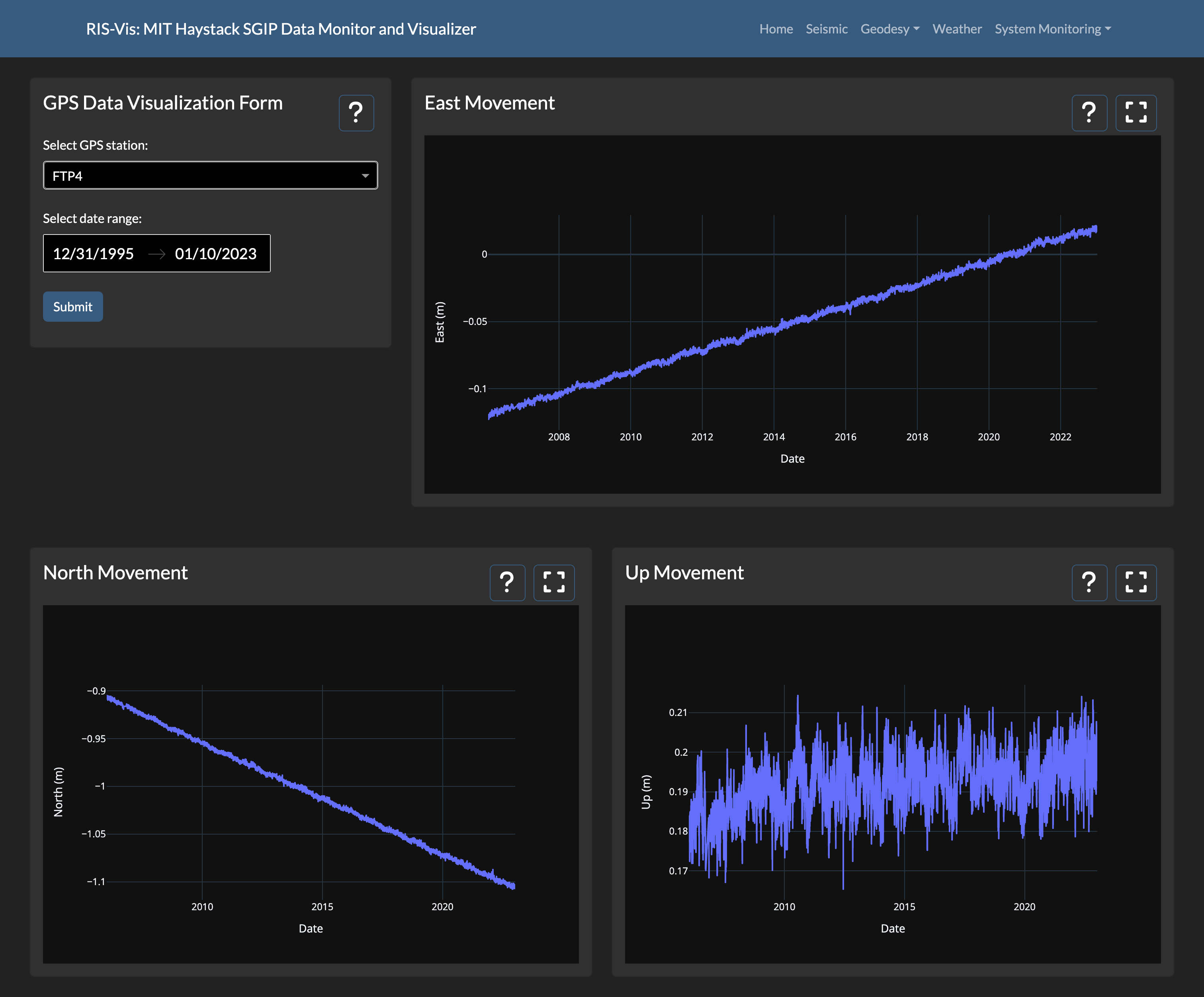}%
    \label{fig:gps}}
    \hfil
\end{figure*}
\begin{figure*}[htbp] \ContinuedFloat 
    \centering 
    \subfloat[]{\includegraphics[width=3.5in]{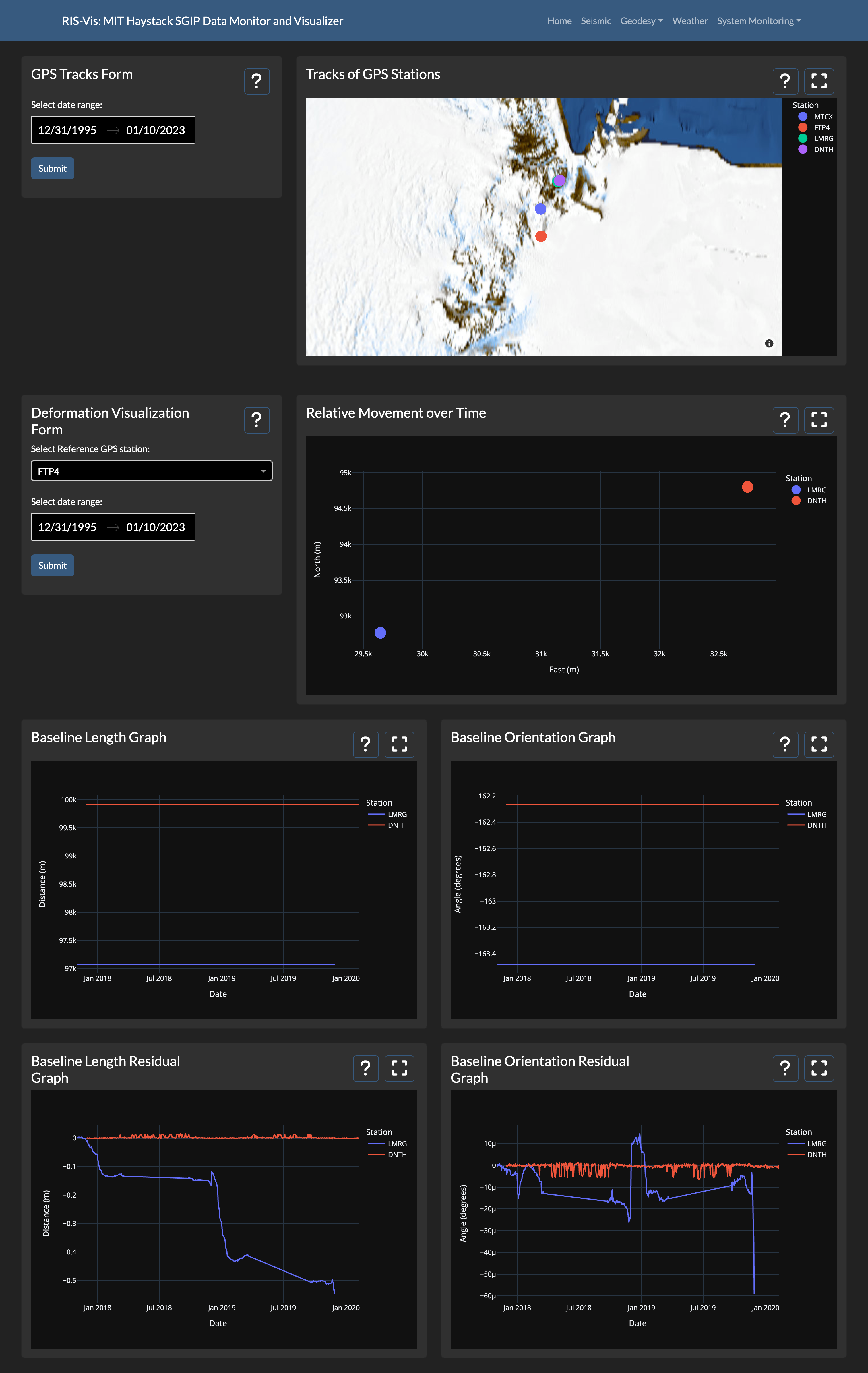}%
    \label{fig:gpsvis}}
    \hfil
    \subfloat[]{\includegraphics[width=3.5in]{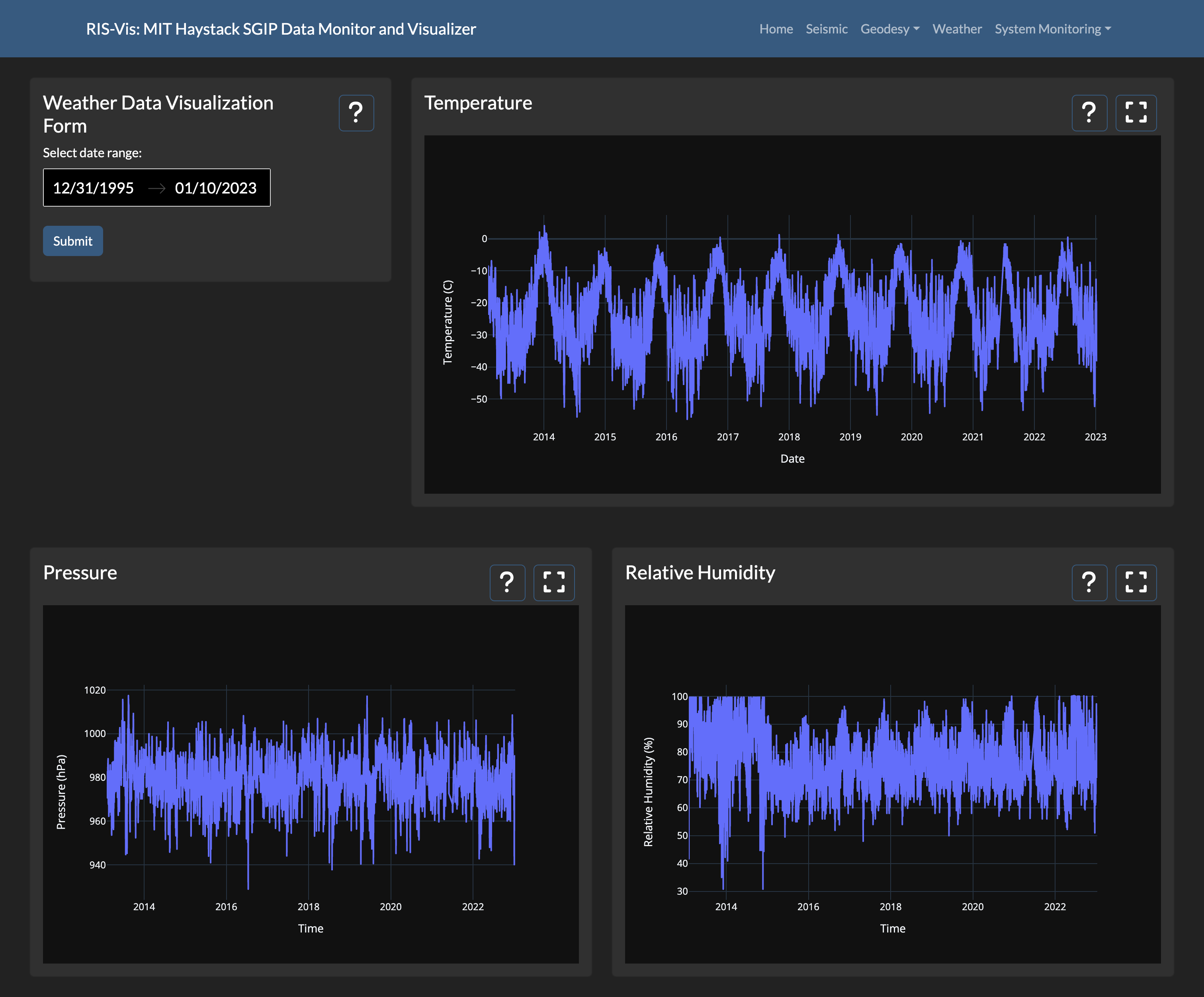}%
    \label{fig:weather}}
    \caption{RIS-Vis front-end views for (a) Seismic, (b) Geodetic East, North, and Up motions, (c) Geodetic deformation, (d) Weather. The Home and System Monitoring pages can be viewed by running RIS-Vis locally.} 
\end{figure*}

\subsubsection{Views}
RIS-Vis contains five different sections as part of its layout, namely, a “Home” page, “Seismic” page, “Geodesy” subsection, “Weather” page, and a “System Monitoring” subsection, which we now describe in turn.
\begin{itemize}
    \item{\textbf{Home page}\\
    The home page of RIS-Vis comprises all real-time visualizations pertaining to SGIP data.
    \begin{list}{}{}
        \item{\textbf{Weather.} The top row of the home page contains current weather conditions as well as a four-day forecast of the Ross Ice Shelf, sourced from the OpenWeatherMap API. Weather data is updated hourly upon refresh of RIS-Vis.}
        \item{\textbf{Spectrogram Visualizations.} The second row of the home page contains spectrogram visualizations of waveform data over the last three days for all seismic stations from which data is collected. Spectrograms are updated every six hours upon refresh of RIS-Vis.}
        \item{\textbf{Log Tables.} When the MIT Haystack Observatory's SGIP instrument is deployed and real-time data is transferred via an Iridium satellite, it is vital for scientists to know what data has been downloaded and transferred to the database. Thus, the bottom row of the home page displays three tables corresponding to each type of data source (seismic, geodetic, weather). The tables outline what the latest files that have been downloaded from each data source are and when they were downloaded.}
    \end{list}
    } 
    \item {\textbf{Seismic page}
    \begin{list}{}{}
        \item{\textbf{Filtering Form.} The form located at the top-left of the seismic page (Figure~\ref{fig:seismic_page}) allows users to filter what seismic data they would like to visualize. Some of the input features include date range, seismic station, and waveform filters (including displacement, velocity, and acceleration).}
        \item{\textbf{Filtered Waveform.} The waveform graph displays seismic waveform data filtered based on the selected option (displacement, velocity, acceleration) in the filtering form.}
        \item{\textbf{Spectrogram.} The spectrogram on the seismic page is developed utilizing the selected waveform data. Spectrograms can be used to detect infragravity waves that have the potential to enlarge crevasses within the ice shelf.}
        \item{\textbf{Power Spectral Density.} The seismic page displays a power spectral density (PSD) visualization that allows researchers to determine the relative amplitude of various frequencies over a certain period of time. PSD visualizations are useful to identify the origin of the response of the ice shelf as ocean swell, infragravity waves, tsunami waves, or other forcings that may have impacted on the ice shelf.}
    \end{list}
    }

    \item {\textbf{Geodetic subsection}
    \begin{list}{}{}
        \item{\textbf{East, North, and Up Movements Page.} The “East, North, and Up Movements” page contains information about the relative east/north/vertical position of GPS stations over time. East positions, measured in eastings, are calculated relative to a reference longitude, while north positions, measured in northings, are calculated relative to the equator. Only “fractional” components of eastings/northing/vertical distances are plotted to represent movement over time at the beginning of the graphs; however, fractional components continue to increase past 1 if necessary with respect to the reference longitude/equator in the Nevada Geodetic Laboratory data utilized to develop RIS-Vis\cite{ref8}. }
        \item{\textbf{Deformation Visualizations page.}  The “Deformation Visualizations” page (Figure~\ref{fig:gpsvis}) contains more detailed visualizations pertaining to GPS station movement over time. The top section of the deformation visualizations page contains an overlay of station positions on a mapbox, displaying station movement over time on a map. The bottom section of the “Deformation Visualizations” Page displays relative station positions (represented by position, distance, and bearing angle) to one selected station. }
    \end{list}
    }

    \item {\textbf{Weather page}
    \begin{list}{}{}
        \item{\textbf{Temperature, Pressure, and Relative Humidity plots.} Based on the date-range selected in the weather form located on the top left of the weather page (Figure~\ref{fig:weather}), line-plots containing temperature, pressure, and relative humidity at the Marilyn Station in the Ross Ice Shelf will be displayed. }
    \end{list}
    }

    \item {\textbf{System Monitoring subsection}\\ 
    The system monitoring subsection will be used to display utility data about the SGIP instrument when deployed. As mentioned in Table 1, as the SGIP is not yet deployed, the system monitoring subsection data contains proxy data.
    \begin{list}{}{}
        \item{\textbf{Gyroscope and Acceleration page.} The gyroscope and acceleration page contains scatter plots displaying information about the gyroscopic and acceleration movements of the instrument in the X, Y, and Z directions.}
        \item{\textbf{Pressure and Humidity page.} The pressure and humidity page displays scatter plots containing information about the pressure and humidity inside the instrument. }
        \item{\textbf{Utility page.} The utility page contains scatter plots that provide information about CPU and memory usage, as well as the amount of free disk space in the instrument.}
        \item{\textbf{Voltage, Current, and Temperature page.} The voltage, current, and temperature page displays scatter plots containing information about incoming and outgoing voltage and current to the battery of the instrument, as well as the temperature inside the instrument. }
    \end{list}
    }
    
\end{itemize}

\subsection{Optimization Decisions}
As RIS-Vis incorporates seismic, geodetic, and weather data spanning a period of several years, optimizations become necessary to be able to display large amounts of data in some scenarios. The optimizations include:
\begin{itemize}
    \item{\textbf{Caching}: A \textbf{Redis Cache} was heavily utilized to optimize the home page of RIS-Vis. As the weather predictions and spectrograms were only updated every hour/six-hours respectively, a time-expiring cache was utilized to reduce unnecessary computations.}
    \item{\textbf{Resampling}: Displaying \textbf{Waveform Data} spanning more than a couple days is a computationally intensive task, as waveform data spanning one day contains 86,400 points if the sampling rate is 1 sample/sec. As Plotly’s web graphing capabilities can currently only handle around 500,000 points, optimizations were necessary to account for larger scatter plots. We utilized the Python \textbf{Plotly Resampler} library to aid in displaying large waveform data; the library downsamples data depending on the zoom-range to make the number of points plotted at any given range feasible.}
    \item{\textbf{Downsampling}: Displaying large \textbf{spectrograms} led to significant speed drawbacks utilizing Plotly’s graphing capabilities. To ensure fast loading times for spectrograms no matter the time range is chosen, the Python library \textbf{Datashader} was utilized to downsample spectrograms to 800 x 300 pixels.}
\end{itemize}

\section{Challenges}

The main challenge encountered in developing RIS-Vis was increasing the efficiency of calculations. Though various optimizations to improve the speed of RIS-Vis were made as detailed above, several computationally-intensive scientific calculations still create delayed loading times. For example, on the “Seismic Page”, calculations for the spectrogram and power spectral density are computationally significant and take loading times as seen in Table~\ref{tab:load_time}.

To solve this issue in the future, a cache could be implemented to mitigate repeated calculations. However, as there are endless possible inputs for the form on the “Seismic Page,” the cache system must be developed to account for the values of various inputs without exponentially increasing the size of the cache unnecessarily.

\begin{table}
\caption{Loading Times for Seismic Graphs\label{tab:load_time}}
\centering
\begin{tabular}{|c|c|}
\hline
Days of Data & Loading Time (s)\\
\hline
1 & ~3\\
\hline
10 & ~10\\
\hline
20 & ~15\\
\hline
30 & ~23\\
\hline
40 & ~30\\
\hline
50 & ~35\\
\hline
\end{tabular}
\end{table}

\section{Summary and Future Work}
Ice shelves are a critical component of the Antarctic cryosphere and climate system. There is a need for a tool that can help scientist to continuously monitor their state and stability. Towards that end, we have developed RIS-Vis, a scalable and interactive visualization dashboard that can be used to monitor seismic, geodetic, and weather data. Although RIS-Vis currently incorporates data from a variety of public repositories, the end goal of RIS-Vis is to utilize data collected from a new instrument called the Seismogeodetic Ice Penetrator (SGIP), currently being built by the MIT Haystack Observatory and soon scheduled to be deployed on the Ross Ice Shelf, Antarctica.

RIS-Vis is currently capable of generating many visualizations that can help scientists ascertain the current state of the Ross Ice Shelf and its response to environmental forcings. For example, by monitoring the presence of infragravity waves utilizing spectrograms, visualizing climate change in the Ross Ice Shelf over time, and tracking how the ice shelf is flowing over time, scientists can readily use past and present-day data to make predictions of possible future scenarios for the stability of the Ross Ice Shelf. 

However, there are still many potential improvements that can be made to further automate the process of visualizing ice shelf stability. For example, machine learning algorithms could be utilized to detect the response of the RIS to infragravity and other ocean waves in a spectrogram; instead of forcing scientists to recognize the presence of a particular type of ocean wave on their own, RIS-Vis could be adapted to notify scientists whenever an ocean wave has been detected by trained convolutional neural network (CNN) models. 

Finally, although the primary purpose of RIS-Vis is to track the health of the Ross Ice Shelf, the same development process utilized to build RIS-Vis could be used to build other general ice shelf monitoring applications (as long as significant data is available). Since RIS-Vis was developed using modular components (to make it feasible to quickly add and/or remove components), the structure of RIS-Vis could easily be modified to better fit the needs of any particular organization, accommodating a variety of scientific goals.

\section{Acknowledgements}
This study was funded by the National Science Foundation (NSF) Research Opportunities for Undergraduates (REU) program at MIT Haystack Observatory award numbers 1950348 and 2243909, and the NSF Office of Polar Programs (OPP) award number 1931131. This work acknowledges the ``Severo Ochoa Centre of Excellence" accreditation (CEX2019-000928-S) funded by AEI 10.13039/501100011033.

\begin{IEEEbiographynophoto}{Aishwarya Chakravarthy} is an undergraduate computer science student at Georgia Institute of Technology at Atlanta, GA, 30332, USA. Her research interests include applied machine learning and human-computer interaction (HCI). Contact her at \href{achakrav6@gatech.edu}{achakrav6@gatech.edu}.
\end{IEEEbiographynophoto}
\begin{IEEEbiographynophoto}{Dhiman Mondal} is a research scientist at the MIT Haystack Observatory at Westford, MA, 01886, USA. His research interests include space geodetic systems, such as Global Navigation Satellite System (GNSS) and Very Long Baseline Interferometry (VLBI), climate systems, and ground deformation. Contact him at \href{dmondal@mit.edu}{dmondal@mit.edu}.
\end{IEEEbiographynophoto}
\begin{IEEEbiographynophoto}{John Barrett} is a radio science software developer at the MIT Haystack Observatory at Westford, MA, 01886, USA. His research interests include writing software with a focus on data processing for VLBI and other scientific applications. Barrett received his Ph.D. in Physics from the Massachusetts Institute of Technology in 2016 while working on the KATRIN neutrino experiment. Contact him at \href{barrettj@mit.edu}{barrettj@mit.edu}.
\end{IEEEbiographynophoto}
\begin{IEEEbiographynophoto}{Chester ``Chet'' Ruszczyk} a research scientist at the MIT Haystack Observatory at Westford, MA, 01886, USA. His research interests include instrumentation development in the field of Very Long Baseline Interferometry (VLBI) for geodetic and astronomy applications (e-VLBI, recorders, digital backends, cable delay measurement systems) along with remote sensor technology, using GNSS, satellite communication, and seismometers for cryosphere applications in both the Arctic and the Antarctic. Ruszczyk received his Ph.D. in Electrical Engineering from Boston University, Boston, Massachusetts. Contact him at \href{chester@mit.edu}{chester@mit.edu}.
\end{IEEEbiographynophoto}
\begin{IEEEbiographynophoto}{Pedro Elosegui} is a research scientist at the MIT Haystack Observatory at Westford, MA, 01886, USA and the Institute of Marine Sciences (ICM) of the Spanish National Research Council (CSIC) at Barcelona, 08003, Spain. His reseearch interests include space geodesy, developing new geodetic technologies, and devising new geoscience applications. Contact him at \href{elosegui@mit.edu}{elosegui@mit.edu}.
\end{IEEEbiographynophoto}

\end{document}